\documentclass[12pt]{article}
\title{Comment about quasi-isotropic solution of Einstein equations near
cosmological singularity}
\author{I.M. Khalatnikov$^{1,2,3}$, A.Yu. Kamenshchik$^{1,2}$ and A.A.
Starobinsky$^{1}$}
\date{}
\begin{document}
\maketitle
\hspace{-5mm}$^1$L.D. Landau Institute for Theoretical Physics of Russian
Academy of Sciences, Kosygin str. 2, 117334, Moscow, Russia\\
$^2$Landau Network - Centro Volta, Villa Olmo, via Cantoni 1, 22100 Como,
Italy\\
$^{3}$Tel Aviv University,
Tel Aviv University, Raymond and Sackler
Faculty of Exact Sciences, School of Physics and Astronomy,
Ramat Aviv, 69978, Israel\\

\begin{abstract}
We generalize for the case of arbitrary hydrodynamical matter the
quasi-isotropic solution of Einstein equations near cosmological singularity,
found by Lifshitz and Khalatnikov in 1960 for the case of the 
radiation-dominated Universe. It is shown that this solution always exists,
but dependence of  terms in the quasi-isotropic expansion acquires a more
complicated form.
\end{abstract}
PACS: 04.20-q; 04.20.Dw\\

In the paper \cite{Lif-Khal} by Lifshitz and Khalatnikov, the quasi-isotropic
solution of the Einstein equations near a cosmological singularity was found
provided the Universe was filled by radiation  with the equation of state $p
= \frac{\varepsilon}{3}$. The metric of this solution was written down in
the synchronous system of reference
\begin{equation}
ds^2 = dt^2 - \gamma_{\alpha\beta}dx^{\alpha}dx^{\beta},
\label{sinchron}
\end{equation}
where spatial metric $\gamma_{\alpha\beta}$ near the singularity has the
form
\begin{equation}
\gamma_{\alpha\beta} = t a_{\alpha\beta} + t^2 b_{\alpha\beta} + \cdots
\label{expansion}
\end{equation}
where $a_{\alpha\beta}$ and $b_{\alpha\beta}$ are functions of spatial
coordinates.  The functions $a_{\alpha\beta}$ are chosen arbitrary, and then
the functions $b_{\alpha\beta}$ and also the energy and velocity
distributions for matter can be expressed through these functions (for
details see \cite{Lif-Khal} and also
\cite{Lif-Khal1,Land-Lif}).

In correspondence with the standard cosmological model of the hot Universe,
it was supposed that the natural equation of state for the matter near the
cosmological singularity is that of radiation: $p=\varepsilon/3$. However, 
nowadays
the situation has changed in connection with the development of inflationary
cosmological models, which as an important ingredient contain inflaton
scalar field or/and other exotic types of matter \cite{inflation}. 
One can add also that the appearance of brane and M theory 
cosmological models \cite{brane} and
the discovery of the cosmic acceleration \cite{cosmic} suggests that the
matter playing essential role on different stages of cosmological evolution
can obey very different equations of state \cite{varun}.
Thus, generalization of the old quasi-isotropic solution of the Einstein
equations near the cosmological singularity can be useful in this new
context. 

In this note we make such a generalization for the equation of state:
\begin{equation}
p = k \varepsilon,
\label{state}
\end{equation}
where $p$ denotes pressure and $\varepsilon$ denotes energy density
\footnote{Actually, the authors have known this generalization for a long 
time but have never published it in detail in regular journals.}.  
The Friedmann isotropic solution near the singularity for such a matter
behaves as
\begin{equation}
a \sim a_0 t^m,
\label{Friedmann}
\end{equation}
where
\begin{equation}
m = \frac{4}{3(1+k)}.
\label{Friedmann1}
\end{equation}
We look for an expression for a spatial metric in the following form:
\begin{equation}
\gamma_{\alpha\beta} = t^m a_{\alpha\beta} + t^n b_{\alpha\beta},
\label{metric}
\end{equation}
where the power index $m$ is given by Eq. (\ref{Friedmann1}). We leave the 
power index $n$ free for some time, requiring only that
\begin{equation}
n > m.
\label{inequality}
\end{equation}
The inverse metric reads
\begin{equation}
\gamma^{\alpha\beta} = \frac{a^{\alpha\beta}}{t^m} -
\frac{b^{\alpha\beta}}{t^{2m-n}},
\label{inverse}
\end{equation}
where $a^{\alpha\beta}$ is defined by the relation
\begin{equation}
a^{\alpha\beta} a_{\beta\gamma} = \delta_{\gamma}^{\alpha}
\label{inverse1}
\end{equation}
while the indices of all the other matrices are lowered and raised  by
$a_{\alpha\beta}$ and $a^{\alpha\beta}$, for example,
\begin{equation}
b_{\beta}^{\alpha} = a^{\alpha\gamma}b_{\gamma\beta}.
\end{equation}
Let us write down also expressions for the extrinsic curvature, its
contractions and its derivatives:
\begin{equation}
\kappa_{\alpha\beta} \equiv \frac{\partial \gamma_{\alpha\beta}}{\partial t}
= mt^{m-1}a_{\alpha\beta} + nt^{n-1}b_{\alpha\beta},
\label{extrinsic}
\end{equation}
\begin{equation}
\kappa_{\alpha}^{\beta} = \frac{m \delta_{\alpha}^{\beta}}{t} +
\frac{(n-m)b_{\alpha}^{\beta}}{t^{m-n+1}},
\label{extrinsic1}
\end{equation}
\begin{equation}
\kappa_{\alpha}^{\alpha} = \frac{3m}{t} + \frac{(n-m)b}{t^{m-n+1}},
\label{extrinsic2}
\end{equation}
\begin{equation}
\frac{\partial\kappa_{\alpha}^{\beta}}{\partial t} =
-\frac{m\delta_{\alpha}^{\beta}}{t^2} -
\frac{(m-n+1)(n-m)b_{\alpha}^{\beta}}{t^{m-n+2}},
\label{extrinsic3}
\end{equation}
\begin{equation}
\frac{\partial\kappa_{\alpha}^{\alpha}}{\partial t} =
-\frac{3m}{t^2} - \frac{(m-n+1)(n-m)b}{t^{m-n+2}},
\label{extrinsic4}
\end{equation}
\begin{equation}
\kappa_{\alpha}^{\beta}\kappa_{\beta}^{\alpha} = \frac{3m^2}{t^2} +
\frac{2m(n-m)b}{t^{m-n+2}}.
\label{extrinsic5}
\end{equation}
We need also an explicit expression for the determinant of the spatial
metric:
\begin{equation}
\gamma \equiv \det \gamma_{\alpha\beta} = t^{3m}(1+t^{n-m}b)\det a,
\label{determin}
\end{equation}
\begin{equation}
\dot{\gamma} \equiv \frac{\partial \gamma}{\partial t} = (3mt^{3m-1} +
b(2m+n)t^{2m+n-1})\det a,
\label{determin1}
\end{equation}
\begin{equation}
\frac{\dot{\gamma}}{\gamma} =
\frac{3m}{t}\left(1+\frac{b(n-m)t^{n-m}}{3m}\right).
\label{determin2}
\end{equation}

Now, using well-known expressions for the components of the Ricci tensor
\cite{Land-Lif}:
\begin{equation}
R_0^0 = -\frac{1}{2} \frac{\partial \kappa_{\alpha}^{\alpha}}{\partial t} -
\frac{1}{4}\kappa_{\alpha}^{\beta}\kappa_{\beta}^{\alpha},
\label{curvature}
\end{equation}
\begin{equation}
R_{\alpha}^0 = \frac{1}{2}(\kappa_{\alpha;\beta}^{\beta} -
\kappa_{\beta;\alpha}^{\beta}),
\label{curvature1}
\end{equation}
\begin{equation}
R_{\alpha}^{\beta} = -P_{\alpha}^{\beta} -\frac{1}{2}\frac{\partial
\kappa_{\alpha}^{\beta}}{\partial t} - \frac{\dot{\gamma}}{4\gamma}
\kappa_{\alpha}^{\beta},
\label{curvature2}
\end{equation}
where $P_{\alpha}^{\beta}$ is a three-dimensional part of the Ricci tensor,
and substituting into Eqs. (\ref{curvature})-(\ref{curvature2}) the
expressions (\ref{extrinsic})-(\ref{determin2}), one get
\begin{equation}
R_0^0 = \frac{3m(2-m)}{4t^2} - \frac{(n-1)(n-m)b}{2t^{m-n+2}},
\label{curvature3}
\end{equation}
\begin{equation}
R_{\alpha}^0 = \frac{n-m}{2t^{m-n+1}}(b_{\alpha;\beta}^{\beta} -
b_{;\alpha}),
\label{curvature4}
\end{equation}
\begin{eqnarray}
&&R_{\alpha}^{\beta} = -\frac{\tilde{P}_{\alpha}^{\beta}}{t^m} +
\frac{m(2-3m)\delta_{\alpha}^{\beta}}{4t^2}\nonumber \\
&&+\frac{(n-m)(2-2n-m)b_{\alpha}^{\beta}}{4t^{m-n+2}} -
\frac{m(n-m)b\delta_{\alpha}^{\beta}}{4t^{m-n+2}}.
\label{curvature5}
\end{eqnarray}

Notice, that in Eq. (\ref{curvature5}) $\tilde{P}_{\alpha}^{\beta}$ denotes
a three-dimensional Ricci tensor constructed by using the metrics
$a_{\alpha\beta}$. The terms in the curvature tensor $P_{\alpha}^{\beta}$,
which are proportional to $\beta_{\alpha\beta}$ have the time dependence
$\sim \frac{1}{t^{2m-n}}$ and are less divergent than the first term in the
right-hand side of Eq. (\ref{curvature5}) provided the condition
(\ref{inequality}) is satisfied.

Now, let us write down the expressions for the components of the
energy-momentum tensor of the perfect fluid
\begin{equation}
T_{ik} = (\varepsilon + p)u_i u_k - p g_{ik},
\label{tensor}
\end{equation}
satisfying the equation of state (\ref{state}). Up to higher-order
corrections, they have the following form :
\begin{equation}
T_0^0 = \varepsilon
\label{tensor1}
\end{equation}
\begin{equation}
T_{\alpha}^0 = \varepsilon(k+1)u_{\alpha},
\label{tensor2}
\end{equation}
\begin{equation}
T_{\alpha}^{\beta} = -k \varepsilon \delta_{\alpha}^{\beta},
\label{tensor3}
\end{equation}
\begin{equation}
T = T_i^i = \varepsilon (1-3k).
\label{tensor4}
\end{equation}
Using the Einstein equations
\begin{equation}
R_i^j = 8\pi G (T_i^j - \frac{1}{2}\delta_i^j T),
\label{Einstein}
\end{equation}
one has from $00$-component of these equations:
\begin{equation}
8\pi G\varepsilon = \frac{1}{3k+1}\left(\frac{3m(2-m)}{2t^2} -
\frac{(n-1)(n-m)b}{t^{m-n+2}}\right),
\label{energy}
\end{equation}
and from $0\alpha$-component of these equations one has
\begin{equation}
u_{\alpha} = \frac{(n-m)(3k+1)(b_{\alpha;\beta}^{\beta} -
b_{;\alpha})t^{n+1-m}}{3m(2-m)(k+1)}.
\label{velocity}
\end{equation}
Now, writing down the spatial components of the Einstein equations, using
the expressions (\ref{tensor3})-(\ref{tensor4})
and  the expression (\ref{energy}) for the energy density $\varepsilon$, 
one get:
\begin{eqnarray}
&&-\frac{\tilde{P}_{\alpha}^{\beta}}{t^m} +
\frac{m(2-3m)\delta_{\alpha}^{\beta}}{4t^2} +
\frac{(n-m)(2-2n-m)b_{\alpha}^{\beta}}{4t^{m-n+2}} \nonumber \\
&&-\frac{m(n-m)b\delta_{\alpha}^{\beta}}{4t^{m-n+2}}
=
\frac{(k-1)\delta_{\alpha}^{\beta}}{3k+1}\left(\frac{3m(2-m)}{2t^2}-\frac{(n
-1)(n-m)b}{t^{m-n+2}}\right).
\label{Einstein1}
\end{eqnarray}

Using the relation (\ref{Friedmann1}) it is easy to check that the terms
proportional to $\frac{1}{t^2}$ in the left- and right-hand sides of Eq.
(\ref{Einstein1}) cancel each other.
On the other hand, the only way to cancel the term
$\frac{\tilde{P}_{\alpha}^{\beta}}{t^m}$ is to require that the terms
proportional to $\frac{1}{t^{m-n+2}}$ behave as the term
$\frac{1}{t^m}$, i.e.
\begin{equation}
n = 2.
\label{secondorder}
\end{equation}
In this case, the condition of cancellation of terms proportional to
$\frac{1}{t^m}$ gives the following expression for the tensor
$b_{\alpha}^{\beta}$:
\begin{equation}
b_{\alpha}^{\beta} = \frac{4\tilde{P}_{\alpha}^{\beta}}{m^2-4}
+ \frac{\tilde{P}\delta_{\alpha}^{\beta}(-3m^2+12m-4)}
{3m(m-3)(m^2-4)}.
\label{b}
\end{equation}
Using the relation (\ref{Friedmann1}) one can rewrite the expression
(\ref{b}) in the following form:
\begin{equation}
b_{\alpha}^{\beta}
= -\frac{9(k+1)^2}{(3k+5)(3k+1)}\left(\tilde{P}_{\alpha}^{\beta}
 + \frac{(3k^2-6k-5)\delta_{\alpha}^{\beta}\tilde{P}}{9k+5}\right).
\label{b1}
\end{equation}

It is easy to see that the Eq. (\ref{b1}) expressing the second-order
correction for the spatial metric (\ref{expansion}) is well defined for all
the types of hydrodynamical matter with $0 \leq k \leq 1$, including the
stiff matter, i.e. a fluid with the equation of state $p = \varepsilon$.

Now, using the relation
\begin{equation}
\tilde{P}_{\alpha}^{\beta ;\alpha} = \frac{1}{2}\tilde{P}_{;\beta},
\label{Bianchi}
\end{equation}
and the formulae (\ref{b1}) and (\ref{velocity}) we arrive to the following
expression for the three-dimensional velocity $u_{\alpha}$:
\begin{equation}
u_{\alpha} = -\frac{27 k (k+1)^3 \tilde{P}_{;\alpha}}{8(3k+5)(9k+5)}
t^{3-\frac{4}{3(k+1)}}.
\label{velocity1}
\end{equation}
Note that the velocity flow is potential. Actually, it can be shown that this
important property remains in all higher orders of perturbative expansion
for the quasi-isotropic solution.
Similarly, the expression for the energy density of matter (\ref{energy})
can be rewritten as
\begin{equation}
8\pi G\varepsilon = \frac{4}{3(k+1)^2t^2} +
\frac{9(k+1)\tilde{P}}{2(9k+5)t^{\frac{4}{3(k+1)}}}.
\label{energy1}
\end{equation}

It is straightforward to check that further terms of perturbative expansion
(\ref{expansion}) for the metric $\gamma_{\alpha\beta}$ 
have the following form:
\begin{equation}
\gamma_{\alpha\beta} = t^m a_{\alpha\beta} + t^2 b_{\alpha\beta} +
t^{2+(2-m)} c_{\alpha\beta}+\cdots.
\label{expansion1}
\end{equation}
Thus, we have seen that this expansion has a curious feature. The order of
its first term $t^m$ is defined by the equation of state of the matter
(\ref{state}), the second order term always has the behavior $\sim t^2$,
while logarithmic distance between orders is equal to $2-m$.

It is easy to understand that the quasi-isotropic expansion does work if and
only if the first term in the right-hand side of Eq. (\ref{metric}) 
is smaller than the second one. Remembering that $n = 2$ for any
value of $m$ and using the equation  (\ref{Friedmann1}), we get the
following restriction on the parameter $k$ from the equation of state
(\ref{state}):
\begin{equation}
k > -\frac{1}{3}.
\label{restr}
\end{equation}
Thus, for the values $k \leq -1/3$ the quasi-isotropic expansion at small
times may not be constructed.
It is interesting to notice, however, that for $k =const < - \frac{1}{3}$,
a quasi-isotropic-like solution arises as a late-time ($t \rightarrow \infty$)
attractor for generic inhomogeneous evolution of space-time  \cite{Star}
(this regime is the power-law inflation \cite{L} actually).
Of course, perturbative expansion is made in inverse powers of $t$ in that 
case. Also, we would like to note that different aspects of
relation between the quasi-isotropic expansion and other approximation 
schemes were considered in detail in the papers \cite{Deruelle}.

This work was supported by RFBR via grants No 02-02-16817 and 00-15-96699.
A.K. is grateful to the CARIPLO Scientific Foundation for the financial
support.


\end{document}